%% file: main.tex
\documentclass[runningheads,envcountsame]{llncs}
\usepackage{graphicx}
\usepackage[inline]{enumitem}
\usepackage{hyperref}
\usepackage{tikz}
\usepackage{amsmath}
\usepackage{algorithm}
\usetikzlibrary{arrows}
\usepackage[noend]{algpseudocode}
\usepackage{comment}
\usepackage{listings}

\usepackage{cite}

\usepackage{booktabs}

\renewcommand\L{{\mathcal L}}
\begin{document}
\title{Distributed and Adversarial Resistant Workflow Execution on the Algorand Blockchain}
\titlerunning{Distributed and Adversarial Resistant Workflow Execution}
\author{Yibin Xu\inst{1}
\and Tijs Slaats\inst{1}
\and Boris D{\"u}dder\inst{1}
\and S{\o}ren Debois\inst{2}
\and Haiqin Wu\inst{1}
}

\authorrunning{Y. Xu \and T. Slaats \and B. D{\"u}dder  \and  S. Debois\and H. Wu}

\institute{
University of Copenhagen, Copenhagen, Denmark\\
\email{\{yx,slaats,boris.d,hw}@di.ku.dk\}\\
\and
IT University of Copenhagen, Copenhagen, Denmark\\
\email{\{debois\}@itu.dk}\\
}

\maketitle    

\begin{abstract}
 
We provide a practical translation from the Dynamic Condition Response (DCR) process modelling language to the Transaction Execution Approval Language (TEAL) used by the Algorand blockchain. Compared to earlier implementations of business process notations on blockchains, particularly Ethereum, the present implementation is four orders of magnitude cheaper. This translation has the following immediate ramifications: (1) It allows decentralised execution of DCR-specified business processes in the absence of expensive intermediaries (lawyers, brokers) or counterparty risk. (2) It provides a possibly helpful high-level language for implementing business processes on Algorand. (3) It demonstrates that despite the strict limitations on Algorand smart contracts, they are powerful enough to encode models of a modern process notation.


\keywords{Applications of blockchain 
\and Smart contracts 
\and Algorand 
\and Inter-institutional collaboration}
\end{abstract}

\input{macros}

\newif\ifjournal
\journalfalse
\newif\ifanonymised
\anonymisedfalse

\section{Introduction}
Blockchain technologies rose to prominence by realising decentralised financial systems and instruments~\cite{buterin2014next,gilad2017algorand}, then branched out into other domains such as supply chain management\ifjournal\ifanonymised ~\cite{saberi2019blockchain}\else ~\cite{saberi2019blockchain,dudder2017timber,COWI21}\fi~and governmental processes\ifanonymised ~\cite{OLNES2017355,8038519}\else ~\cite{OLNES2017355,8038519,10.1007/978-3-030-51280-4_36}\fi. \else~\cite{saberi2019blockchain}. \fi
The main draw of blockchains is their ability to securely capture and track the ownership of resources~\cite{zakhary2019global}, e.g., digital cash, real estate, and produce. Smart contracts\ifjournal~\cite{szabo1997formalizing}~\else~\cite{buterin2014next}~\fi have added a second dimension to these use cases by allowing blockchains to control the valid movement of resources. 
\setlength{\parskip}{0em} 

This development has drawn the interest of the Business Process Management (BPM) community\ifanonymised ~\cite{mendling2018blockchains}\else ~\cite{mendling2018blockchains}\fi, to which
smart contracts harbor the promise of integrity-protected decentralised automation of process execution. In this community, a process is commonly defined as \emph{a structured, measured set of activities designed to produce a specific output for a particular customer or market}~\cite{Davenport:1993:PIR:171556}. In practice, e.g., products being traded and shipped in a supply chain and the treatment of a patient in a hospital, or a loan application process within a bank. The latter two of these examples are knowledge-intensive processes.  
A key approach to formalising knowledge-intensive
processes is that of \emph{declarative process notations}~\cite{Pesic2007},
which expresses the constraints a process must obey, as opposed to the exact
sequencing of admissible activity executions, akin to the difference between an LTL formula
and a B{\"u}chi automaton. In practice, declarative models tend to be more concise, and for processes subject to rules and regulations, easier to relate to those rules and regulations. 

While process notations have been encoded into Solidity~\cite{buterin2014next}), these are plagued by high costs and relatively low performance due to congestion of the Ethereum network and high gas prices\ifanonymised ~\cite{anon} \else ~\cite{madsen2018collaboration}\fi. Given the cost of transactions on the Ethereum blockchain in January 2022, creation of a declarative business process contract would cost roughly \$350 while the execution of a single event or activity would cost \$25.

In the current paper we address this weakness by exploring an encoding from the declarative Dynamic Condition Response (DCR) Graphs process notation to TEAL~\cite{gilad2017algorand} contracts running on the Algorand blockchain. Transactions on Algorand are cheap, with a current cost, on the 14th of January 2022, of \$0.00136, and offer transaction finality in under 5 seconds. 
This encoding is not trivial however, as the efficiency and low cost of TEAL contracts carry limitations to the memory space and number of operations as a trade-off.

\paragraph{The contributions} of this paper are:
\begin{enumerate}
    \item We show how DCR Graphs can be efficiently stored in the limited memory space provided by TEAL and through pseudo-code show how their run-time semantics can be encoded without exceeding the operation limit;
    \item we analyse the costs of storing and running DCR smart contracts on the Algorand blockchain based on the number of unique activities involved in the process;
    \item we provide a prototype implementation of the encoding running on the Algorand testnet;
    \item we discuss possible future extensions to the encoding that will allow capturing more complex and rich process descriptions.
\end{enumerate}

In section 2, we proceed to discuss future work. In section 3, we shortly describe the primary attributes of the Algorand network. Section 4 introduces Dynamic Condition Response (DCR) graphs. In section 5, we show how the semantics of DCR Graphs can be encoded as smart contracts on the Algorand blockchain. Section 6 provides a financial analysis, show the maximum cost associated with our approach and how it related to earlier attempts at encoding DCR Graphs in Solidity. Finally Section 7 concludes and discusses future work.

\section{Related work}
Previous approaches towards process-aware blockchains \ifjournal \cite{tran2018lorikeet,lopez2019caterpillar,ladleif2019modeling,klinger2020blockchain} \else \cite{tran2018lorikeet,lopez2019caterpillar,ladleif2019modeling,klinger2020blockchain} \fi have focused on providing translations from process models into existing smart contracts languages, particularly, by translating flow-based BPMN diagrams to Solidity. 
A recent work \cite{lopez2019interpreted} proposed to reduce the cost of redeployment of the smart contracts when changing the process model by a specially designed interpreter of BPMN process models based on dynamic data structures. \cite{lopez2020controlled} presented a model for dynamic binding of actors to roles in collaborative processes and an associated binding policy specification language. 
We differ from these works by first of all, taking a declarative approach to process modelling and second of all developing a native smart contract language for processes that is directly embedded in the blockchain.

Inspired by institutional grammars, \cite{frantz2016institutions} proposed a high-level declarative language that focuses on business contracts, however, no implementation is provided. 
A high-level vision of the business artifact paradigm towards modelling business processes on a distributed ledger was given in \cite{hull2016towards}. 
\cite{10.1007/978-3-030-11641-5_29} proposed a lean architecture enabling lightweight and full-featured on-chain implementations of a decentralised process execution system.
\ifjournal In \ifanonymised ~\cite{madsen2018collaboration}\else \cite{madsen2018collaboration}\fi, the authors mapped DCR Graphs to Solidity contracts but showed that efficiency and costs are open challenges.\fi

\section{Algorand blockchain}
Algorand\cite{gilad2017algorand} is a late-generation blockchain with a series of features, including high scalability and a fork-free consensus protocol based on Proof-of-Stake. Its smart contract layer (ASC1) aims to reduce the security risk of smart contracts, and adopts a non-Turing complete programming model, which natively supports transactional atomic sets and user-defined assets. These characteristics make it an intriguing smart contract platform to study.

A smart contract language called TEAL \cite{gilad2017algorand} is used in Algorand. TEAL is a bytecode-based stack language and is processed by the Algorand Virtual Machine (AVM), with an official programming interface for Python (called PyTeal). In addition to standard arithmetic-logical operators, TEAL also includes operators for calculating and indexing all transactions in the current atomic group, as well as IDs and fields for accessing them. When launching a transaction involving a script, the user can specify a series of parameters. The script includes cryptographic operators that calculate the hash value and verify the signature.

Applications are stateful smart contracts created with Algorand. They are given an Application ID when they are launched. Application Transactions are used to communicate with these contracts. The primary Application Transaction provides additional data that the stateful smart contract's TEAL code can pass and process.

Per transaction call, any application can check the global state of up to two other smart contracts. This is accomplished by including the application IDs of the additional stateful smart contracts in the transaction call to the stateful smart contract. This is known as the Application Array in TEAL. Currently, the developer must know how many additional applications are expected to be sent into the contract call before writing the smart contract code in TEAL.

\begin{figure}
    \centering
    \includegraphics[width=0.9\textwidth]{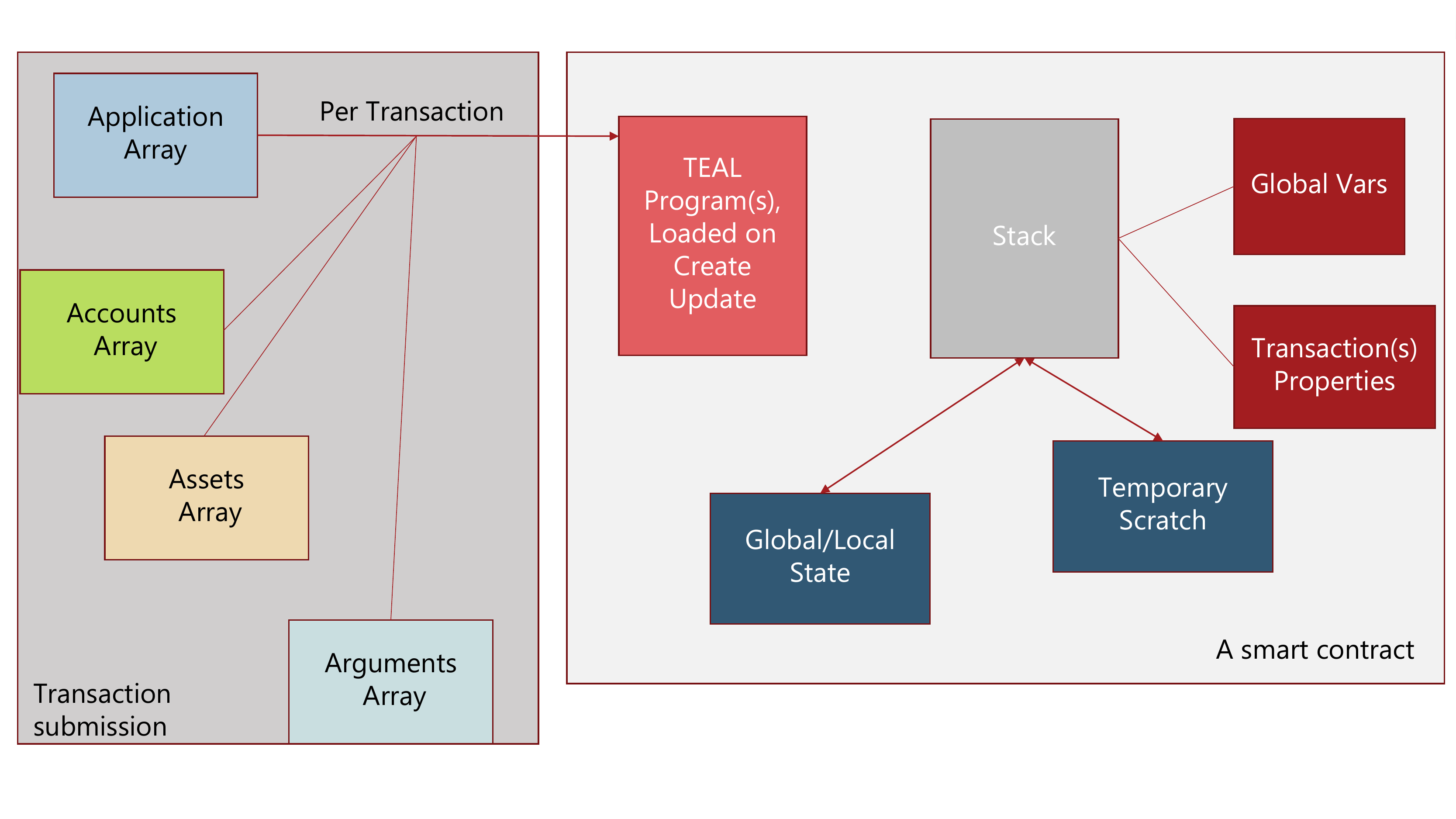}
    \caption{Stateful Smart Contract Transaction Call Architecture.}
    \label{fig:algorand}
\end{figure}

Figure \ref{fig:algorand} shows the architecture of the stateful smart contract in Algorand. Each transaction has an Application array, which indicates what smart contract (up to two smart contracts) the transaction will be sent to; an Accounts Array (up to four accounts), which indicates what accounts have opt-in to the smart contract; an Assets Array (up to two assets), which indicates the assets that will be sent to the smart contract; an Arguments Array (up to 255 arguments), which indicates the arguments passed to the smart contract. The maximum length of the stack and scratch space is 1000 and 255 respectively.

\subsection{Limitations}\label{3.1}
For each smart contract, we have the following limitations:
\begin{enumerate}
    \item 64 Key/Value pairs in the global state.
    \item 64 Key/Value pairs in total in the local state of the accounts. There can be four accounts opted in to the smart contract, each of it has 16 pairs. 
    \item Max key + Value length=128 bytes
    \item A stateless smart contract only returns the result of the execution but will not store states in the blockchain. A stateful smart contract can create and update the states stored in the blockchain.
    \item The program (a smart contract consists of an approval program and a clean program) costs no more than 20000 operations in stateless mode. It allows at most 700 operations for each the approval and clear program of a stateful contract. In other words, it allows 700 operations for each execution of the stateful contract. Each operation costs 1, some cryptographic operations are more costly but not used in our prototypical implementation.       
\end{enumerate}
Note that the pairs in the local state of the accounts are read-only for other people. People do not need to Opt-in to the smart contract in order to execute the smart contract.
\section{DCR Graph}  
\label{sec:dcr}
DCR Graphs~\cite{DBLP:journals/corr/abs-1110-4161} are a formal declarative notation for describing processes. The
base notation focuses on control-flow, i.e., the allowed sequencing of
activities. The nodes of a DCR Graph are the executable elements, called
events, which can be labelled by a labelling function. The labelling function
allows multiple events to share the same label, thereby allowing process
activities to occur more than once in a graph, under different constraints
depending on their context.

The state of a graph is described by a \emph{marking}, indicating for each
event whether it (1) has been previously executed, (2) is currently pending,
and (3) is currently included. The evolution of the graph is described by its edges, the relations between events.
Through the relations, an event can constrain another event or have an effect
on it. There are two possible constraints: the condition ($\conditionrel$)
captures that an event can not be executed unless another event has been
executed some time (not necessarily immediately) before it, the milestone
($\milestonerel$) captures that an event cannot be executed while another
event is pending. There are three effect relations: the exclusion
($\excluderel$) removes an event from the process and disables any
constraints it may place on other events, the inclusion relation
($\includerel$) includes an event back into the process, re-enabling any
constraints it may have had, and the response relation makes another event
pending ($\responserel$).
Pending events are obligations and must be satisfied by either executing or excluding them before a process can be considered to be in an accepting state.

\begin{definition}
\label{def:dcr}
A \emph{DCR Graph} is a tuple $(E,M,L, \labellingfun, \conditionrel,\responserel,\milestonerel,\includerel,\excluderel)$, where
\begin{itemize}
    \item $E$ 
    is the set of events
    \item $M =(\Ex,\Re,\In)
    \in {\cal{P}}(E) \times {\cal{P}}(E) \times {\cal{P}}(E)$
    is the \emph{marking} of the graph 
    \item $L$ is the set of \emph{labels}
    \item $\labellingfun:E \to L$ is the \emph{labelling function}
    \item $\phi\subseteq E\times E$ for $\phi\in \reltypes$ are respectively the condition, response, milestone, inclusion, and exclusion \emph{relations} between events
    \end{itemize}
 \end{definition}

For DCR Graph $G$ with events $E$ 
and event $e\in E$, we write
$\eventrelationto{e}{\conditionrel}$ for the set $\{e'\in E\mid  e'\conditionrel e\}$, write $\eventrelationfrom{e}{\responserel}$ for the set $\{e'\in E\mid e \responserel e'\}$ and similarly for $\eventrelationfrom{e}{\includerel}$, $\eventrelationfrom{e}{\excluderel}$ and $\eventrelationto{e}{\milestonerel}$.


An event of a DCR graph is \emph{enabled} when (a) it is included, (b) there are no included conditions that have not been executed, and (c) there are no pending and included milestones.

\begin{definition}[Enabled events]
    \label{def:enabled} 
    Let $G=(E,M,L, \labellingfun, \conditionrel,\responserel,\milestonerel,\includerel,\excluderel)$ be a DCR Graph, with marking
    $M = (\Ex,\Re,\In)$.
    An event $e \in E$ is \emph{enabled}, written 
    $\enables G e$, iff
    (a) $e \in \In{}$ and (b) $\In{}\cap
    \eventrelationto{e}{\conditionrel} \subseteq \Ex$ and (c) $(\Re{}\cap\In)\cap
    \eventrelationto{e}{\milestonerel} = \emptyset$. 
  \end{definition}

If an event is enabled then it can be executed. Executing an event $e$ updates the marking of the graph by (a) adding it to the set of executed events, (b) removing it from the set of pending events and adding its responses $\eventrelationfrom{e}{\responserel}$ to the set of pending events, and (c) respectively removing its exclusions $\eventrelationfrom{e}{\excluderel}$ from and adding its inclusions $\eventrelationfrom{e}{\includerel}$ to the set of included events.

\begin{definition}[Execution]
    \label{def:execution}
    Let $G=(E,M,L, \labellingfun, \conditionrel,\responserel,\milestonerel,\includerel,\excluderel)$ be a DCR Graph, with marking $M = (\Ex,\Re,\In)$. When $\enables G e$, the result of executing $e$, written $\execute G e$ is a new DCR Graph $G'$ with the same events, labels, labelling function, and relations, but a new marking $M'= (\Ex',\Re',\In')$, where (a) $\Ex' = \Ex \cup \{e\}$ (b) $\Re' = (\Re \backslash \{e\}) \cup \eventrelationfrom{e}{\responserel}$, and (c) $\In' = (\In\backslash \eventrelationfrom{e}{\excluderel}) \cup
    \eventrelationfrom{e}{\includerel}$.
\end{definition}





We define the language of a DCR Graph as all finite and infinite sequences of such executions, where all pending responses are eventually executed or excluded.

\begin{definition}[Language of a DCR Graph]
    \label{def:runs} 
    Let $G=(E,M,L, \labellingfun, \conditionrel,\responserel,\milestonerel,\includerel,\excluderel)$ be a DCR Graph. A \emph{run} of $G$ is a finite or infinite
    sequence of events $e_0, e_1,\ldots$ such that $\enables{G_i}{e_i}$, $\execute{G_i}{e_i} = G_{i+1}$, and $G_0 = G$.
    We call a run \emph{accepting} iff for each $G_i$ with marking $M_i = (\Ex_i, \Re_i, \In_i)$ and $e \in \Re_i \cap \In_i$ there exists a $j \geq i$ such that $e_j=e$ or $e \not\in \Re_j \cap \In_j$.

    The \emph{language} $\lang {G}\subseteq L^{\infty}$ of $G$ is the set of finite and infinite sequences of labels $l_0l_1\cdots$ such that there is an accepting run $e_0, e_1,\ldots$ where $\labelling{e_i}=l_i$.
  \end{definition}
  \let\L\lang

\begin{figure}[tbh]
    \includegraphics[width=\textwidth]{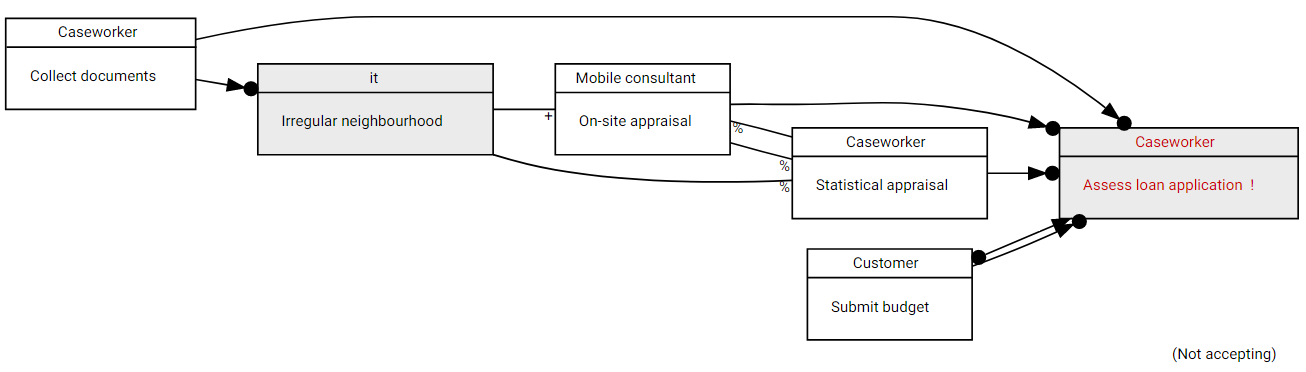}
    \caption{DCR Graph of a mortgage application process adapted from\ifanonymised ~\cite{anon}\else ~\cite{DCRBPM2016}\fi.}
    \label{fig:example}
\end{figure}
  
As an example, Fig.~\ref{fig:example} shows a simplified version of a loan application process encountered in industry\ifanonymised ~\cite{anon} \else ~\cite{DCRBPM2016} \fi modelled as a DCR Graph. The labels of the events contain not only the name of the activity, but also the roles who are allowed to execute them. The loan application should always be assessed by the case worker, shown by the red text and exclamation mark, which denote that the event is an \emph{initial response}.
To reach this goal, the case worker must first collect documents and the customer must submit a budget, shown by the condition relations from these two events. In addition, a statistical or on-site appraisal must have been performed. Both are a condition to assess loan application, but they also mutually exclude each other, meaning that if one is executed, the other is excluded and will not block other events from executing. 
Submit budget also has a response relation towards the assessment, meaning that a loan application must always be assessed (again) after the customer submits a (new) budget. Finally IT may determine that the neighbourhood of the property requires an on-site appraisal. It then excludes the statistical appraisal event and includes the on-site appraisal event, which will re-enable on-site as a condition for the assessment, even if it was previously excluded by a statistical appraisal.

In the initial marking, irregular neighbourhood and assess loan application are blocked as having unsatisfied conditions. Other events are enabled as they are included and have no blocking conditions or milestones. The graph is in a non-accepting state as the assess loan application is included and a pending response.

Executing Collect documents and Submit budget will mark these events as executed. Doing a Statistical appraisal will mark itself as executed and exclude On-site appraisal, meaning that we can execute Assess loan application, which will remove the pending response and bring the graph into an accepting state.

Note that we can still execute Submit budget if new information is provided by the customer, which requires Assess loan application to be executed again.  
  
\section{Distributed DCR Graphs as Algorand smart contracts}
In this section, we transform the DCR Graph into a stateful smart contract in TEAL. We eliminate the labels of the events and only keep the relationships and the IDs of the events. This design is for maintaining the anonymity of the DCR Graphs in the blockchain and saving space for more events.

For each DCR Graph, we maintain three global key/value pairs:
\begin{itemize}
    \item $GC$, which records the address of the graph creator as a Byte32 String.
    \item $MK$, which indicates the marking of the graph as a Byte16 String;
    \item $TEN$, the total event number as an unsigned 64-bit integer. \footnote{Note that integer in TEAL is automatically a uint 64 integer. }\end{itemize}
 Each four bits of $MK$ represents the status of an event, with the first bit describing if the event is included or excluded. The structure of the status: 
 \begin{itemize}
     \item Excluded: $(xxx0)_2$;
     \item Included: $(xxx1)_2$;
     \item Pending: $(xx1x)_2$;
     \item executed: $(x1xx)_2$.
 \end{itemize} where $x$ represents either $1$ or $0$. The number $i, i\in[0,TEN\times 5)$ bit refers to a status of the number $int(i/5)$ event. Only $CG$ can add events or add the relationships between the events.

We maintain two key/value pairs $E$ and $E\_links$ for each event $E$. The key/value pair $E$ indicates the account address which can execute $E$; $E\_links$ indicates the links between $E$ and other events (which event's status needs to be changed after the execution of $E$ and which events are preventing $E$ from execution) as a Byte32 String. Each five bits of the Value represents the links of an event. The structure of the links: 
\begin{itemize}
    \item Include: $(xxxx1)_2$;
    \item Exclude: $(xxx1x)_2$;
    \item Milestone: $(xx1xx)_2$;
    \item Condition: $(x1xxx)_2$;
    \item Response: $(1xxxx)_2$.
\end{itemize} 
Include, Exclude, Response are out-links, meaning that after execution of $E$, the relevant event will be included, excluded or pended. Milestone and Condition are in-links, meaning $E$ may not be executed if the relevant event is pended or have not executed. For example, given two events $A$ and $B$ with relations $A \conditionrel B$ and $B \excluderel A$, $B\_links$ will indicate both $A\conditionrel B$ and $B \excluderel A$. $B\_links=(0100000010)_2$ assuming $A$ indexed $1$ and $B$ indexed $2$.

Given the limitations discussed in Section~\ref{3.1}, our approach can have 61 events in maximum because we have 128 pairs in total and we use two key/value pairs for each event and three pairs for the graph.

\begin{figure}
    \centering
    \includegraphics[width=0.9\textwidth]{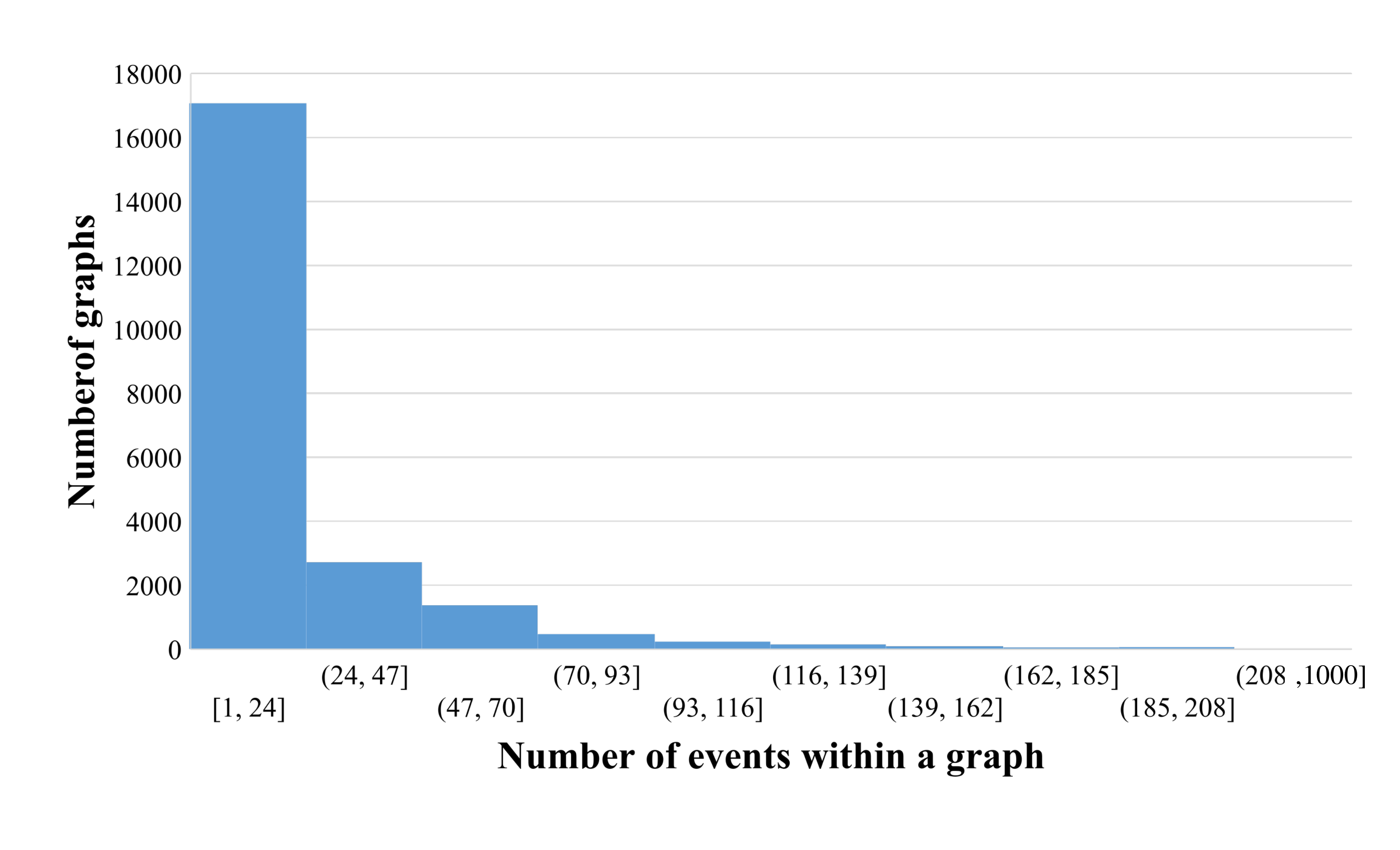}
    \caption{The statistic from \url{https://www.dcrgraphs.net/}.}
    \label{fig:statistic}
\end{figure}

We were provided with a breakdown of a database containing 22787 DCR models created by academic and commercial users; we report a summary in Figure~\ref{fig:statistic}. Note that the statistics show that the average number of events within a graph generated in the site is 23 and 92.5\% of the graphs have an event number below 61. Therefore, the 61 event limit appears to be a promising start for practical usage. 

\begin{figure}
    \centering
    \includegraphics[width=1\textwidth]{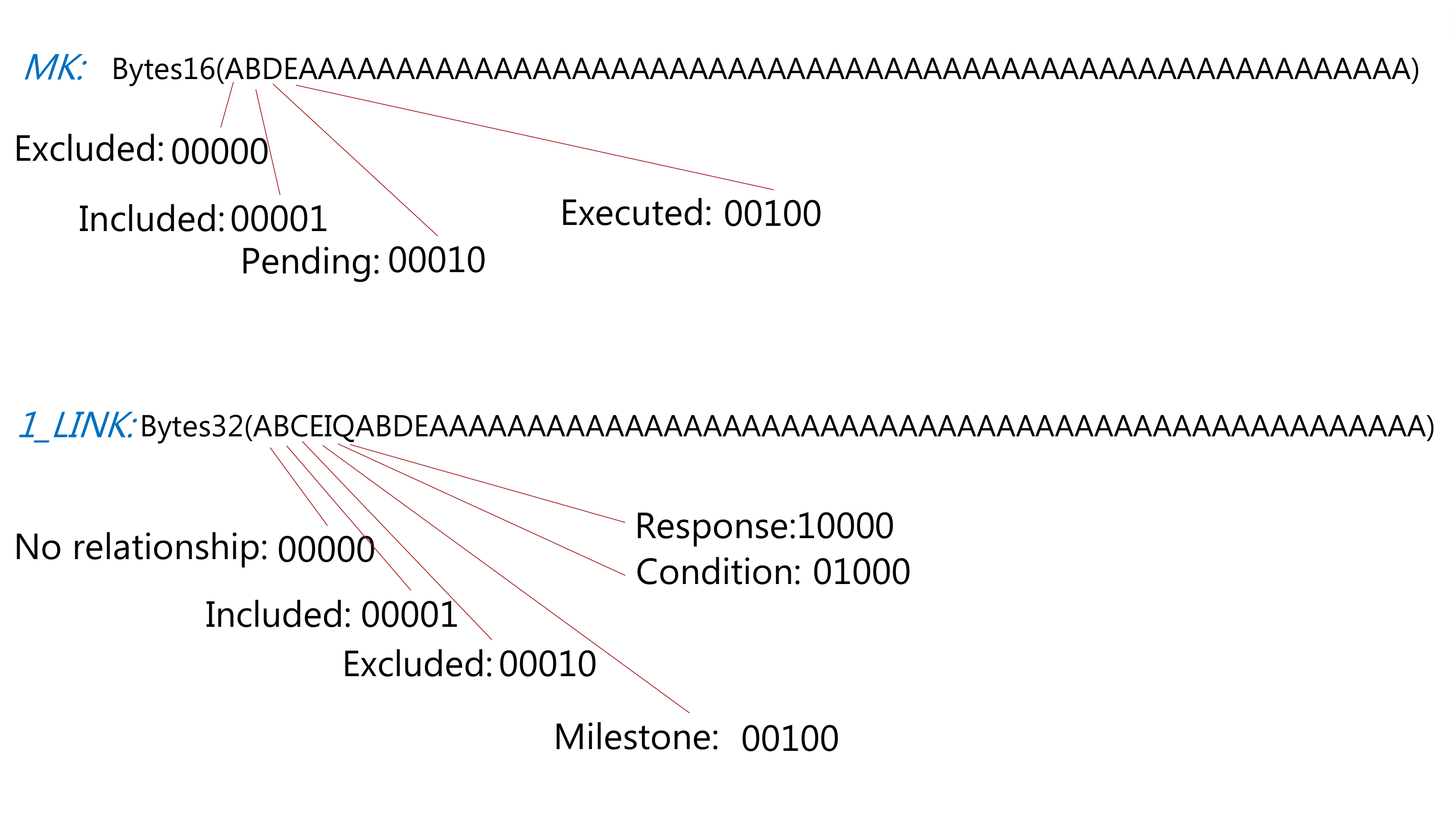}
    \caption{The structures of the Key/Value pairs.}
    \label{fig:mk}
\end{figure}

Figure~\ref{fig:mk} shows an example of the architecture, and $1\_LINK$ indicates the link K/V pair of event 1. 

Algorithms \ref{alg1}, \ref{alg2}, \ref{alg3}, and \ref{alg4} show the pseudo-code of the operations of adding an event, adding an relationship, executing an event, and updating the status of the events, respectively. In Algorithm \ref{alg1}, we add an event by creating two key/value pairs representing the executor and the links to other events. In Algorithm \ref{alg2}, we add an relationship between two events by updating their links. The executor of an event can execute the event in Algorithm \ref{alg3}, the Algorithm will check in-links of the event to see if it is executable and then update $MK$ via the out-links. In Algorithm \ref{alg4}, $CG$ can update the status of an event.

In the codes, we are not using any operations that require a cost of more than 1. Therefore the four Algorithms are all within 700 operations when there are 61 events in maximum.

An example implementation corresponding to the graph shown in Figure \ref{fig:example} is in the Algorand testnet, APP-ID:59565714. The link to the Github repository for the source code is \url{https://github.com/XU-YIBIN/DCR-Algorand}.

\begin{algorithm}
	\caption{Add an event} 
	\label{alg1}
	\begin{algorithmic}[1]
	\State Global states: $GC:Graph\ creator$, $MK:Marking$, $TEN: Total\ event\ number$
	\Procedure{Add an event}{$TXS:Transaction\ sender$, $EC:executor$}     
	\Require {$TXS==GC$ and $TEN<61$}
	\State $TEN \gets TEN+1$
	\State Set a K/V pair $A$: 
	\newline $ _{ }$ \textbf{Key:} $TEN\_links$,  $ _{ }$ \textbf{Value:} $Byte32(Null)$
	\State Set a Scratch value $total\_ops$, //scratch value is run-time value.
	\State $ _{ }$ $total\_ops \gets (TEN-1)\times 2 +3 +1$ \newline // because we have used three key/value pairs at the beginning and we are currently adding one more. 
	\If{ $total\_ops>64$}
	\State Set $A$ as a local state in the account $[(total\_ops-64)/16 -1]$
	\Else $ _{ }$ Set $A$ as a global state.
	\EndIf
	\State $ _{ }$ $total\_ops \gets total\_ops+1$
	\State Set a K/V pair $B$:
	\newline  $ _{ }$ \textbf{key:} $TEN$, \textbf{Value:}$EC$ 
	\If{ $total\_ops>64$}
	\State Set $B$ as a local state in the account $[(total\_ops-64)/16 -1]$
	\Else $ _{ }$ Set $B$ as a global state.
	\EndIf
	\EndProcedure
	\end{algorithmic} 
\end{algorithm}

\begin{algorithm}
	\caption{Add a relationship} 
	\label{alg2}
	\begin{algorithmic}[1]
	\State Global states: $GC:Graph\ creator$, $MK:Marking$, $TEN: Total\ event\ number$
	\Procedure{Add a relationship}{$TXS:Transaction\ sender$, $E1:Event1\ ID$, $E2:Event2\ ID$, $RT:Relationship\ Type$}
	\Require{$TXS==GC$}
	\State Get a K/V pair ($E1\_link$,$A$). \newline //This pair may be from the global state or the local state, which was set when adding the event (using key:$TEN\_links$).
	\State Get a K/V pair ($E2\_link$,$B$). \newline //This pair may be from the global state or the local state.
	\State Set a scratch value $k$ depanding on RT (include$\to$ 0, exclude $\to$ 1 , Milestone $\to$ 2, Condition $\to$ 3, Response $\to$ 4).
	\If {$k=2$ or $k=3$}
	\State Set the $(E1-1)\times 5+k$-bit of $B$ to 1.
	\Else \State Set the $(E2-1)\times 5+k$-bit of $A$ to 1.
	\EndIf
	\State Update K/V pairs ($E1\_link$,A) and ($E2\_link$,B) using updated $A$ and $B$.
	\EndProcedure
	\end{algorithmic} 
\end{algorithm}

\begin{algorithm}
	\caption{Execute an event} 
	\label{alg3}
	\begin{algorithmic}[1]
	\State Global states: $GC:Graph\ creator$, $MK:Marking$, $TEN: Total\ event\ number$
	\Procedure{Execute an event}{$TXS:Transaction\ sender$, $E1:Event1\ ID$}
	\State Get a K/V pair($E1$,$A$). //This pair may from the global state or the local state.
    \Require {$A==TXS$}
	\State Get a K/V pair ($E1\_link$,$B$). \newline //This pair may be from the global state or the local state.
	\For {i=0 to $TXN\times 5$}
	\State Set a scratch value $C\_ID$, $C\_ID\gets int(i / 5)+1$. \newline //$C_ID$ refers to the current event ID.
	\State Set a scratch value $k$, $k \gets i\ mod\ 5$.
	\If {$k$-st bit of $B.Value$=1}
	\If{k=2} // The event $C\_ID$ milestone $E1$.
	\State If the $(C\_ID-1)\times 4+1$-st bit of $MK$ is 1, then return false.
	\ElsIf{k=3} //The event $C\_ID$ condition $E1$.
	\State If the $(C\_ID-1)\times 4+2$-st bit of $MK$ is 0, then return false.
	\EndIf
	\EndIf
	\EndFor
	\For {i=0 to $TXN\times 5$}
	\State Set a scratch value $C\_ID$, $C\_ID\gets int(i / 5)+1$. //$C_ID$ refers to the current event ID.
	\State Set a scratch value $k$, $k \gets i\ mod\ 5$.
	\If {$k$-st bit of $B$ is 1}
	\If{k=0} // The event $C\_ID$ should be included.
	\State Set the $C\_ID\times 4+0$-bit of $MK$ as 1.
	\ElsIf{k=1} //The event $C\_ID$ should be excluded.
	\State Set the $C\_ID\times 4+0$-bit of $MK$ as 0.
	\ElsIf{k=4} //The event $C\_ID$ should be pended.
	\State Set the $C\_ID\times 4+1$-bit of $MK$ as 1.	
	\EndIf
	\EndIf
	\EndFor
	\State Set $(E1-1)\times 4+1$-bit of $MK$ as 0 //cancel the pending status.
	\State Set $(E1-1)\times 4+2$-bit of $MK$ as 1 //update the status as executed.
	\EndProcedure
	\end{algorithmic} 
\end{algorithm}
\begin{algorithm}
	\caption{Update the status of the event} 
	\label{alg4}
	\begin{algorithmic}[1]
	\State Global states: $GC:Graph\ creator$, $MK:Marking$, $TEN: Total\ event\ number$
	\Procedure{Update Status}{$TXS:Transaction\ sender$, $E1:Event1\ ID$, $S:Status$}
    \Require{$TXS==GC$}
	\If {S=``include''}
	\State Set $(E1-1)\times 4$-bit of $MK$ as 1.
	\ElsIf {S=``exclude''}
	\State Set $(E1-1)\times 4$-bit of $MK$ as 0.
	\ElsIf{S=``pend''}
	\State Set $((E1-1)\times 4+1)$-bit of $MK$ as 1.
	\EndIf
	\EndProcedure
	\end{algorithmic} 
\end{algorithm}

\section{Financial Analysis}
When a smart contract is deployed to the Algorand blockchain, it is given an app ID, which is a unique identifier. Furthermore, each smart contract has its own Algorand address, which is created from this unique ID. The address allows the smart contract to function as an escrow account. In order for the smart contract to run, there must be at least $Escrow_{overall}$ amount of microAlgos inside the smart contract, otherwise the transaction fails automatically.

\begin{equation}
    \begin{aligned}
    Escrow_{global}=100000 \times (1+ExtraProgramPages) \\+(25000+3500)\times schema.NumUint\\+(25000+25000)\times schema.NumByteSlice.
\end{aligned}
\end{equation} where $Schema.NumUint$ refers to the global integer key-value pairs (the value size is $UInt64bits$); $Schema.NumByteSlice$ refers to the global string key-value pairs. $ExtraProgramPages$ is only needed when the compiled program exceeds $2KB$, which we do not require. 

The operation opt-in an account to the smart contract requires:
\begin{equation}  \begin{aligned}
Escrow_{local}=100000 +(25000+3500)\times schema.NumUint\\+(25000+25000)\times schema.NumByteSlice.\end{aligned}
\end{equation} $schema.NumUint$ and $schema.NumByteSlice$ refers to the local states.

We use one global integer key-value pair (the number of events), all other key-value pairs are String key-value pairs. $ExtraProgramPages=0$. Then, 
\begin{equation}
    Escrow_{global}=100000+28500+50000\times \min(TSN\times 2+2,63).
\end{equation}
\begin{equation}
\begin{aligned}
    Escrow_{local\_{i,i\in[0,ceil((TSN\times 2 +2 -63)/16))}}\\=100000+50000\times \left (\min\left (TSN\times 2+2-63-(i-1)\times 16,16\right ) \right).
\end{aligned}
\end{equation} 
\begin{equation}
    Escrow_{overall}=Escrow_{global}+ Escrow_{local\_{i,i\in[0,ceil((TSN\times 2 +2 -63)/16))}}.
\end{equation}
When TSN=61,
\begin{equation}
   Escrow_{overall}=Escrow_{global}+ Escrow_{local}\times 4=3278500+3450000=6728500.
\end{equation}

\begin{figure}
    \centering
    \includegraphics[width=\textwidth]{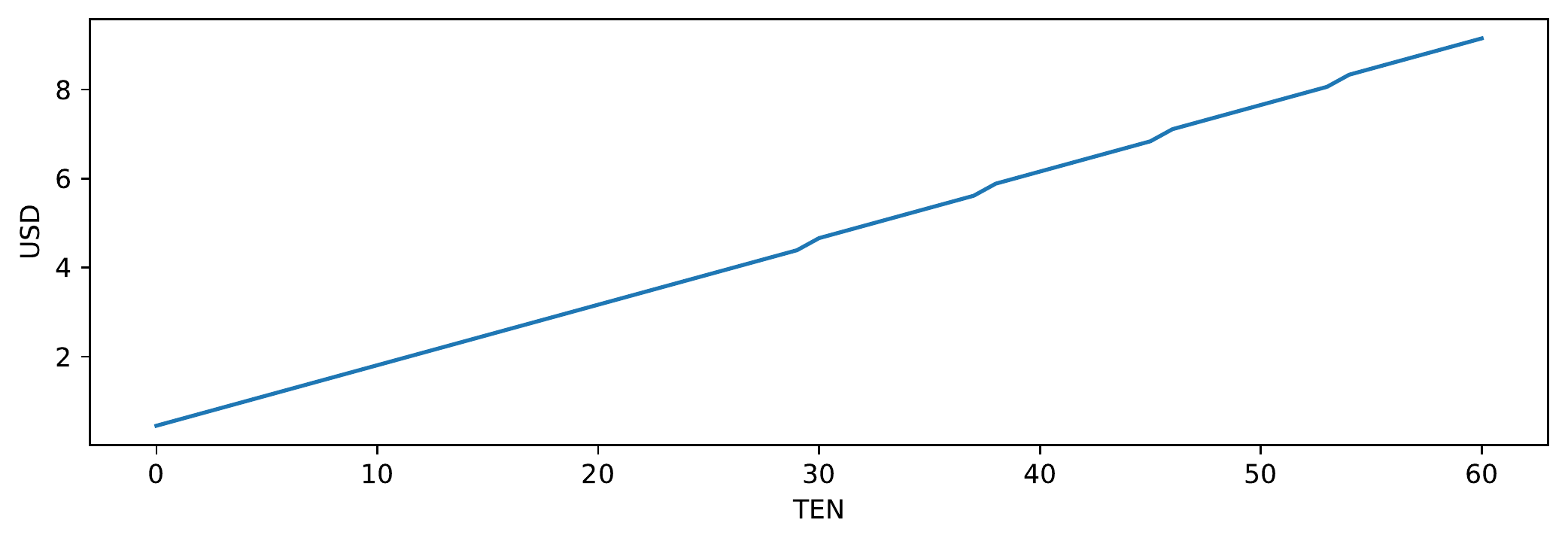}
    \caption{The relationship between the events and the USD escrowed.}
    \label{fig:TEN_USD}
\end{figure}

As of January 14th, 2022, 1000000 microAlgos are worth $\$1.36$. Therefore, the maximum amount of Algo locked for deploying a DCR Graph has a value of $\$9.35$. Figure~\ref{fig:TEN_USD} show the relationship between the number of events in a contract and the amount of USD locked. Note that the escrow is locked in the account that starts the contract and the remaining of it is released when the smart contract is closed. There is a fee for executing the smart contract.

The fees for executing the smart contract is paid by the escrow account linked to the smart contract or can be set to be paid by the executor. Each execution below 1kB costs a fixed 1,000 microAlgos or 0.001 Algos. Larger transactions use a fee-per-byte ratio. One can also choose to use the fee-per-byte ratio to increase the probability of getting included into a new block, however at the current usage levels of the network this is unnecessary.
For our implementation, both contract creation and event execution transactions remain below the 1kB limit.
In addition to the locked escrow, the creation fee for a smart contract is the same as a regular transaction, however since the DCR Graph is dynamically constructed through $addEvent$, $addRelation$, and $updateStatus$ calls to the smart contract, creating the graph will require $1+E+R+S$ transactions, where $E$ is the number of events, $R$ is the number of relations, and $S$ is the number of status updates that need to be made to set the marking of the events of the contract to their initial state.
A comparison of the costs for contract creation and event execution in Algorand and Ethereum is shown in Table~\ref{tab:comparison}. We calculated these numbers based on the example graph used in~\cite{madsen2018collaboration} which contains 5 events, contains 11 relations, and requires 3 status changes to the marking~\footnote{~\cite{madsen2018collaboration} does not provide a generalised calculation of gas costs that can be used for a more thorough comparison.}. We observe a decrease in price by four orders of magnitude for both event execution and contract creation. If one includes the escrow, then contract creation is two orders of magnitude cheaper.

Finally, Algorand provides transaction finality in under 5 seconds, compared to approximately 3 minutes for Ethereum. While the latter is acceptable for many practical business processes, this is a notable improvement for more time-critical scenarios.
\\

\begin{table}[]
\centering
\begin{tabular}{@{}cccc@{}}
\toprule
                                                                                     & Algorand                                                        & Ethereum                                                          & USD Ratio \\ \midrule
\begin{tabular}[c]{@{}c@{}}Contract creation cost\\ (excluding escrow)\end{tabular}  & \begin{tabular}[c]{@{}c@{}}0.02 Algo \\ \$0.02720\end{tabular}  & \begin{tabular}[c]{@{}c@{}}717,709 gas\\ \$349.88314\end{tabular} & 17494     \\
                                                                                     &                                                                 &                                                                   &           \\
\begin{tabular}[c]{@{}c@{}}Contract creation cost \\ (including escrow)\end{tabular} & \begin{tabular}[c]{@{}c@{}}0.7485 Algo\\ \$1.01796\end{tabular} & \begin{tabular}[c]{@{}c@{}}717,709 gas\\ \$349.88314\end{tabular} & 467       \\
                                                                                     &                                                                 &                                                                   &           \\
Event execution                                                                      & \begin{tabular}[c]{@{}c@{}}0.001 Algo \\ \$0.00136\end{tabular} & \begin{tabular}[c]{@{}c@{}}54,496 gas\\ \$26.56690\end{tabular}   & 19534     \\ \bottomrule
\end{tabular}
\caption{Costs and escrow for DCR contract creation and event execution on Algorand and Ethereum based on the example used in~\cite{madsen2018collaboration}. USD prices based on exchange rates on the 14th of January 2022. The calculated dollar cost for Ethereum transactions use a gwei/gas ratio of 150. Event execution cost in Ethereum is given as the mean of the 5 executions reported in~\cite{madsen2018collaboration}.}
\label{tab:comparison}
\end{table}






\section{Conclusion}
In this paper, we demonstrated how business processes can be executed on the Algorand blockchain through a translation from the declarative DCR process modelling language to TEAL smart contracts. We provided precise calculations of limitations on the size of process models and the cost of their execution.
We showed that execution on the Algorand blockchain cuts costs by four orders of magnitude when compared to earlier implementations on Ethereum, bringing the use of public blockchains for business process execution back in the realm of reasonable possibilities.
We implemented a prototype that demonstrates the feasibility of our approach and allows for future extensions.

\paragraph{In future work}, we intend to extend the prototype, lift current limitations and implement more advanced features of the DCR language. In particular, we will extend the current 61 event limit by creating multiple linked smart contracts that can read each others global states. This operation is supported in TEAL by indicating multiple smart contracts in the Application Array of the transactions.

Currently our implementation describes an event by only using an event ID. This is to preserve privacy and save memory space. When the graph is extended by using multiple smart contracts, we may use space to store more information on the events such as their name and a description. 

DCR Graphs support various advanced features such as notions of (logical) time, data-constraints, replication, and more advanced assignments between events, roles, and users~\cite{madsen2018collaboration}, we plan to add these to our encoding in the future which will allow for the description and execution of more complex processes.

\bibliographystyle{splncs04}
\bibliography{main}

\end{document}

%% file: macros.tex
\newif\iflong
\longtrue
\newcommand{\singledcrg}{\text{{DCR Graph}}}
\newcommand{\singleddcrg}{\text{{DDCR Graph}}}

\newcommand{\DCRDefFM}{\ensuremath{(\events,\marking, \conditionrel, \responserel, 
 \includerel,\excluderel, 
\actions,\lab)}}
\newcommand{\DCR}{DCR\xspace}
\newcommand{\DCRR}{DCR$^*$\xspace}
\newcommand{\DCRL}{DCR$^\nu$\xspace}
\newcommand{\DCRB}{DCR$^!$\xspace}
\newcommand{\interact}[3]{\ensuremath{(#1, #2 \rightarrow #3)}}
\newcommand{\rref}{\sqsubseteq}
\renewcommand\t{\ensuremath{\mathsf{t}}}
\newcommand\f{\ensuremath{\mathsf{f}}}

\let\act=\textsf
\newcommand{\events}{\ensuremath{\mathsf{E}}}
\newcommand{\marking}{\ensuremath{\mathsf{M}}}
\newcommand{\actions}{\ensuremath{\mathsf{A}}}
\newcommand{\Roles}{\ensuremath{\mathsf{R}}}
\newcommand{\Actions}{\ensuremath{\mathsf{L}}}
\newcommand{\interactions}{\ensuremath{\mathsf{IA}}}
\newcommand{\binteractions}{\ensuremath{\mathsf{BIA}}}
\newcommand{\lab}{\ensuremath{l}}
\newcommand{\conditionrel}{\ensuremath{\mathrel{\rightarrow\!\!\bullet}}}
\newcommand{\responserel}{\ensuremath{\mathrel{\bullet\!\!\rightarrow}}}
\newcommand{\milestonerel}{\ensuremath{\mathrel{\rightarrow\!\!\diamond}}}
\newcommand{\includerel}{\ensuremath{\mathrel{\rightarrow\!\!+}}}
\newcommand{\excluderel}{\ensuremath{\mathrel{\rightarrow\!\!\%}}}
\newcommand{\init}[1]{\mathsf{Initiator}(#1)}
\newcommand{\responserelop}{\ensuremath{\mathrel{\leftarrow\!\!\bullet}}}
\newcommand{\milestonerelop}{\ensuremath{\mathrel{\rightarrow\!\!\diamond}}}
\newcommand{\includerelop}{\ensuremath{\mathrel{+\!\!\leftarrow}}}
\newcommand{\excluderelop}{\ensuremath{\mathrel{\%\!\!\leftarrow}}}

\newcommand{\responses}{\ensuremath{\mathsf{Re}}}
\newcommand{\executed}{\ensuremath{\mathsf{Ex}}}
\newcommand{\included}{\ensuremath{\mathsf{In}}}

\newcommand{\markingset}{\ensuremath{\mathcal{M}}}
\newcommand{\graphs}{\ensuremath{\mathcal{G}}}
\newcommand{\power}[1]{\ensuremath{{\cal{P}}(#1)}}
\newcommand{\pgraphs}{\ensuremath{{\cal{P}}}}
\newcommand{\inp}[1]{\ensuremath{\triangleleft #1}}
\newcommand{\outp}[1]{\ensuremath{#1\triangleright}}
\newcommand{\eventrelationto}[2]{\ensuremath{\mathord{(#2\!#1)}}}
\newcommand{\eventrelationfrom}[2]{\ensuremath{\mathord{(#1\!#2)}}}

\newcommand{\genrel}{\ensuremath{\mathrel{\rightarrow}}}
\newcommand{\genrelto}[1]{\genrel\!#1}
\newcommand{\genrelfrom}[1]{#1\!\genrel}

\def\L{\mathsf{L}}

\def\lEx{\mathsf{L_{Ex}}}
\def\lRe{\mathsf{L_{Re}}}
\def\lIn{\mathsf{L_{In}}}
\def\dom{\mathsf{dom}}
\newcommand{\reltypes}{\ensuremath{\{\conditionrel,\responserel,\milestonerel,\includerel,\excluderel\}}}
\def\M{\mathsf{M}}

\def\E{\mathsf{\%}}
\def\I{\mathsf{+}}
\def\R{\mathsf{!}}



\newcommand{\id}{\ensuremath{\mathsf{id}}}
\newcommand{\ppevents}{\delta}
\newcommand{\img}[1]{\ensuremath{\mathsf{img}(#1)}}
\newcommand{\execute}[2]{\mathsf{execute}(#1, #2)}
\newcommand{\undefined}{\ensuremath{\mathsf{undefined}}}
\newcommand{\nonlocal}[1]{\ensuremath{\mathsf{nonlocal}(#1)}}

\newcommand{\projection}[2]{{#1}\mathord|_{#2}}
\newcommand{\lift}[2]{{#1}\mathord|^{#2}}

\newcommand\Ex{\mathsf{Ex}}
\renewcommand\Re{\mathsf{Re}}
\newcommand\In{\mathsf{In}}

\newcommand\e[1]{\ensuremath{\mathsf{#1}}}
\newcommand\ev{\mathsf{E}}
\newcommand\fe{\mathsf{fe}}
\newcommand\iface{\mathsf I}
\newcommand\defs{\mathsf S}
\renewcommand\mark{\mathsf M}
\newcommand\rels{\mathsf R}

\newcommand\std{\ensuremath{(\ev, \rels, \mark)}}

\newcommand\local{\mathsf L}

\newcommand{\dcrgraphinline}[7][]{%
\begin{wrapfigure}[#7]{#5}{#6}%
#1%
\centering%
\includegraphics[scale=0.5]{#2}%
\vskip-0.35cm%
\caption{#3#4}%
\vskip-0.5cm%
\end{wrapfigure}%
}

\newcommand{\dcrgraph}[4][]{%
	\begin{figure}[htb]%
	\centering%
    #1\includegraphics[scale=0.5]{#2}%
    \vskip-0.35cm%
	\caption{#3#4}%
    \vskip-0.5cm
	\end{figure}%
}
\newcommand{\enables}[2]{#2\in{\mathsf{enabled}(#1)}}

\newcommand{\transition}[3]{{#1}\xrightarrow{#2}{#3}}

\newcommand{\runs}[1]{\mathop{\mathsf{runs}}(#1)}
\newcommand{\traces}[1]{\mathop{\mathsf{traces}}(#1)}
\newcommand{\lang}[1]{\mathop{\mathsf{lang}}({#1})}
\newcommand{\mustlang}[1]{\mathop{\mathsf{mustlang}}({#1})}
\newcommand{\alphabet}[1]{\mathop{\mathsf{alph}}({#1})}
\newcommand{\dcrzero}{\mathbf{0}}

\newcommand{\iproj}[1]{\projection{#1}{\iota}}
\newcommand{\ilift}[1]{\lift{#1}{\iota}}

\def\L{\mathsf{L}}

\def\lEx{\mathsf{L_{Ex}}}
\def\lRe{\mathsf{L_{Re}}}
\def\lIn{\mathsf{L_{In}}}
\def\dom{\mathsf{dom}}
\def\refines{\mathsf{\,\sqsubseteq\,}}
\def\notRefines{\mathsf{\,\not\sqsubseteq\,}}
\def\M{\mathsf{M}}

\def\E{\mathsf{\%}}
\def\I{\mathsf{+}}
\def\R{\mathsf{!}}

\newcommand{\pev}[1]{{#1}}
\newcommand{\ptresp}[3]{#2\overset{\responserelop}{#3} #1}
\newcommand{\ptcond}[3]{#1\overset{\conditionrel}{#3} #2}

\newcommand{\presp}[2]{#2\responserelop #1}
\newcommand{\pcond}[2]{#1\conditionrel #2}
\newcommand{\pmile}[2]{#1\milestonerel #2}
\newcommand{\pincl}[2]{#2\includerelop #1}
\newcommand{\pexcl}[2]{#2\excluderelop #1}
\newcommand{\pany}[2]{#2 \mathrel{\cal{R}} #1}

\newcommand{\pzero}{0}
\newcommand{\ppar}[2]{#1 \parallel #2}

\newcommand{\psub}[2]{!#1.#2}
\newcommand{\pnewact}[2]{(\nu {#1})\;{#2}}
\newcommand{\pnew}[3]{\pnewact {{#1}:{#2}} {#3}}

\newcommand{\mterm}[1]{{\mathsf{t}}(#1)}
\newcommand{\mmark}[1]{{\mathsf{m}}(#1)}

\newcommand{\marke}[4]{{#1}:({#2}, {#3}, {#4})}
\newcommand{\unknown}{\_}

\newcommand{\process}[2]{[{#1}]\;{#2}}
\newcommand{\action}[2]{{#1}\cdot{#2}}
\newcommand{\hideeffect}[2]{{#1}\mathord\setminus{#2}}
\newcommand{\responsesfun}{\rho}
\newcommand{\lts}[1]{{\mathcal{T}}({#1})}

\def\rulename#1{[\text{\sc #1}]}
\def\ltag#1{\rulename{#1}\quad}
\def\rtag#1{\quad\rulename{#1}}
\newcommand{\Bools}{{\cal B}}
\newcommand{\Events}{{\cal E}}
\newcommand{\Labels}{{\cal L}}
\newcommand{\labelling}[1]{{\ell}({#1})}
\newcommand{\labellingfun}{\ell}
\newcommand{\grey}[1]{\fcolorbox{gray}{gray}{#1}}
\newcommand*{\defeq}{\stackrel{\text{def}}{=}}

\newcommand{\iinsn}[1]{\text{\texttt{#1}}}
\newcommand{\iinc}[2]{\iinsn{inc}(#1,#2)}
\newcommand{\idecjz}[3]{\iinsn{decjz}(#1,#2,#3)}
\newcommand{\ihalt}{\iinsn{halt}}

\newcommand{\insn}[3]{\mathsf{insn}({#1},{#2},{#3})}

\newcommand{\einc}[1]{\iinsn{inc}^{#1}}
\newcommand{\edecjz}[1]{\iinsn{decjz}^{#1}}
\newcommand{\edecjn}[1]{\iinsn{decjn}^{#1}}
\newcommand{\ehalt}{\iinsn{halt}}
\newcommand{\NLOGSPACE}{\textsc{nlogspace}\xspace}
\newcommand{\NPHARD}{\textsc{np-hard}\xspace}
\newcommand{\NPCOMP}{\textsc{np-complete}\xspace}
\newcommand{\EXPTIME}{\textsc{exptime}\xspace}
\newcommand{\NP}{\textsc{np}\xspace}
\newcommand{\encode}[1]{\llbracket{#1}\rrbracket}

\newcommand{\transitionruleinner}[4]{\process {#4} {#2} \vdash {#1}:#3}
\newcommand{\transitionrule}[6]{ \process {#6} {#2} \vdash #1: (#3,#4,#5) }

\newcommand\deadline{\e{deadline}}
\newcommand\recv{\e{recv}}
\newcommand\round{\e{round}}
\newcommand\bm{\e{bm}}
\newcommand\approve{\e{approve}}
\newcommand\reject{\e{reject}}

\renewenvironment{comment}{\par\smallskip\color{red}\noindent\ignorespaces}{\par\smallskip}
\newcommand{\debois}[1]{\par\smallskip\noindent{\small\llap{\textsf{{debois:}~~}}{\textsl{#1}}}\par\smallskip}

\newcommand{\tijs}[1]{\par\smallskip\noindent{\small\llap{\textsf{{tijs:}~~}}{\textsl{#1}}}\par\smallskip}

%% file: main.bbl
\begin{thebibliography}{10}
\providecommand{\url}[1]{\texttt{#1}}
\providecommand{\urlprefix}{URL }
\providecommand{\doi}[1]{https://doi.org/#1}

\bibitem{buterin2014next}
Buterin, V., et~al.: A next-generation smart contract and decentralized
  application platform. white paper  (2014)

\bibitem{Davenport:1993:PIR:171556}
Davenport, T.H.: Process Innovation: Reengineering Work Through Information
  Technology. Harvard Business School Press, Boston, MA, USA (1993)

\bibitem{DCRBPM2016}
Debois, S., Hildebrandt, T., Slaats, T.: Concurrency and asynchrony in
  declarative workflows. In: BPM 2016. LNCS, vol.~9253. Springer, Cham (2016)

\bibitem{frantz2016institutions}
Frantz, C.K., Nowostawski, M.: From institutions to code: Towards automated
  generation of smart contracts. In: FAS*W. pp. 210--215. IEEE (2016)

\bibitem{gilad2017algorand}
Gilad, Y., Hemo, R., Micali, S., Vlachos, G., Zeldovich, N.: Algorand: Scaling
  byzantine agreements for cryptocurrencies. In: SOSP 2017. pp. 51--68 (2017)

\bibitem{DBLP:journals/corr/abs-1110-4161}
Hildebrandt, T.T., Mukkamala, R.R.: Declarative event-based workflow as
  distributed dynamic condition response graphs. In: Honda, K., Mycroft, A.
  (eds.) PLACES 2010. {EPTCS}, vol.~69, pp. 59--73 (2010)

\bibitem{hull2016towards}
Hull, R., Batra, V.S., Chen, Y.M., Deutsch, A., Heath~III, F.F.T., Vianu, V.:
  Towards a shared ledger business collaboration language based on data-aware
  processes. In: ICSOC. pp. 18--36. Springer (2016)

\bibitem{klinger2020blockchain}
Klinger, P., Bodendorf, F.: Blockchain-based cross-organizational execution
  framework for dynamic integration of process collaborations. In: WI (2020)

\bibitem{ladleif2019modeling}
Ladleif, J., Weske, M., Weber, I.: Modeling and enforcing blockchain-based
  choreographies. In: BPM. pp. 69--85. Springer (2019)

\bibitem{lopez2019interpreted}
L{\'o}pez-Pintado, O., Dumas, M., Garc{\'\i}a-Ba{\~n}uelos, L., Weber, I.:
  Interpreted execution of business process models on blockchain. In: EDOC. pp.
  206--215. IEEE (2019)

\bibitem{lopez2020controlled}
L{\'o}pez-Pintado, O., Dumas, M., Garc{\'\i}a-Ba{\~n}uelos, L., Weber, I.:
  Controlled flexibility in blockchain-based collaborative business processes.
  Information Systems p. 101622 (2020)

\bibitem{lopez2019caterpillar}
L{\'o}pez-Pintado, O., Garc{\'\i}a-Ba{\~n}uelos, L., Dumas, M., Weber, I.,
  Ponomarev, A.: Caterpillar: a business process execution engine on the
  ethereum blockchain. SPE  \textbf{49}(7),  1162--1193 (2019)

\bibitem{madsen2018collaboration}
Madsen, M.F., Gaub, M., H{\o}gnason, T., Kirkbro, M.E., Slaats, T., Debois, S.:
  Collaboration among adversaries: distributed workflow execution on a
  blockchain. In: SCFAB 2018 (2018)

\bibitem{mendling2018blockchains}
Mendling, J., Weber, I., Aalst, W.V.D., Brocke, J.V., Cabanillas, C., Daniel,
  F., Debois, S., Ciccio, C.D., Dumas, M., Dustdar, S., et~al.: Blockchains for
  business process management-challenges and opportunities. ACM TMIS
  \textbf{9}(1),  1--16 (2018)

\bibitem{Pesic2007}
Pesic, M., Schonenberg, H., van~der Aalst, W.M.: {DECLARE: Full Support for
  Loosely-Structured Processes}. In: EDOC 2007. pp. 287--287. IEEE (oct 2007)

\bibitem{saberi2019blockchain}
Saberi, S., Kouhizadeh, M., Sarkis, J., Shen, L.: Blockchain technology and its
  relationships to sustainable supply chain management. International Journal
  of Production Research  \textbf{57}(7),  2117--2135 (2019)

\bibitem{10.1007/978-3-030-11641-5_29}
Sturm, C., Szalanczi, J., Sch{\"o}nig, S., Jablonski, S.: A lean architecture
  for blockchain based decentralized process execution. In: Daniel, F., Sheng,
  Q.Z., Motahari, H. (eds.) BPM. pp. 361--373. Springer (2019)

\bibitem{tran2018lorikeet}
Tran, A.B., Lu, Q., Weber, I.: Lorikeet: A model-driven engineering tool for
  blockchain-based business process execution and asset management. In: BPM.
  pp. 56--60 (2018)

\bibitem{zakhary2019global}
Zakhary, V., Amiri, M.J., Maiyya, S., Agrawal, D., Abbadi, A.E.: Towards global
  asset management in blockchain systems (2019)

\end{thebibliography}
